\pdfoutput=1

\documentclass[11pt]{article}

\usepackage{ACL2023}

\usepackage{times}
\usepackage{latexsym}

\usepackage[T1]{fontenc}
\usepackage{amsmath}
\usepackage{amssymb}
\usepackage{subfigure}
\usepackage{graphicx}
\usepackage{subfigure}
\usepackage{multirow}
\usepackage{hyperref} 

\usepackage[utf8]{inputenc}

\usepackage{microtype}

\usepackage{inconsolata}

\title{ATRI: Mitigating Multilingual Audio Text Retrieval Inconsistencies by Reducing Data Distribution Errors}

\author{Yuguo Yin\textsuperscript{1}, \ Yuxin Xie\textsuperscript{1}, \ Wenyuan Yang\textsuperscript{2}, \ Dongchao Yang\textsuperscript{3}, \ Jinghan Ru\textsuperscript{1},\\ \ \textbf{Xianwei Zhuang\textsuperscript{1}, \ Liming Liang\textsuperscript{1}, \ Yuexian Zou\textsuperscript{1}}\thanks{* Yuexian Zou is the corresponding author.}\\
  \textsuperscript{1}Guangdong Provincial Key Laboratory of Ultra High Definition Immersive\\ Media Technology, Shenzhen Graduate School, Peking University,\\
  \textsuperscript{2}School of Cyber Science and Technology, Sun Yat-sen University,\\  
  \textsuperscript{3} The Chinese University of Hong Kong \\
  \texttt{ygyin@stu.pku.edu.cn}, \texttt{zouyx@.pku.edu.cn}\\
  }

\begin{document}
\maketitle
\begin{abstract}
Multilingual audio-text retrieval (ML-ATR) is a challenging task that aims to retrieve audio clips or multilingual texts from databases. However, existing ML-ATR schemes suffer from inconsistencies for instance similarity matching across languages. 
To address the inconsistency issue in multilingual audio-text retrieval, we first identify two intuitive factors that contribute to inconsistency: misalignment between audio and multilingual text embeddings, and error propagation in model optimization. By systematically analyzing these factors, we derive theoretical weight error upper bounds for quantifying their effects and find that the main source of inconsistency is the data distribution error during training. This finding motivates our solution to reduce data distribution errors.
We propose a consistent ML-ATR scheme using 1-to-k contrastive learning and audio-English co-anchor contrastive learning, aiming to mitigate the negative impact of data distribution error on recall and consistency in ML-ATR. Experimental results on the translated AudioCaps and Clotho datasets show that our scheme achieves state-of-the-art performance on recall and consistency metrics for eight mainstream languages, including English. 
Our code will be available at \href{https://github.com/ATRI-ACL/ATRI-ACL}{https://github.com/ATRI-ACL/ATRI-ACL}.

\end{abstract}

\section{Introduction}
In an audio-text retrieval (ATR) task, the system searches for matching audio clips or text captions in a database based on cross-modality queries \cite{zhu2024cacophony,zhuang2024kdpror}. With the convergence of audio and text, ATR techniques have seen significant advancements in recent years and are widely applied in content retrieval and multimedia information retrieval. However, most existing ATR systems are designed for monolingual retrieval, and research on multilingual audio-text retrieval (ML-ATR) remains limited \cite{yan2024bridging}. The shift to ML-ATR brings new challenges, particularly in dealing with high multilingual recall and ensuring the cross-lingual consistency \cite{nie2024improving} of multilingual retrieval results.

\begin{figure}[htbp]
    \centering
    \includegraphics[scale=0.24]{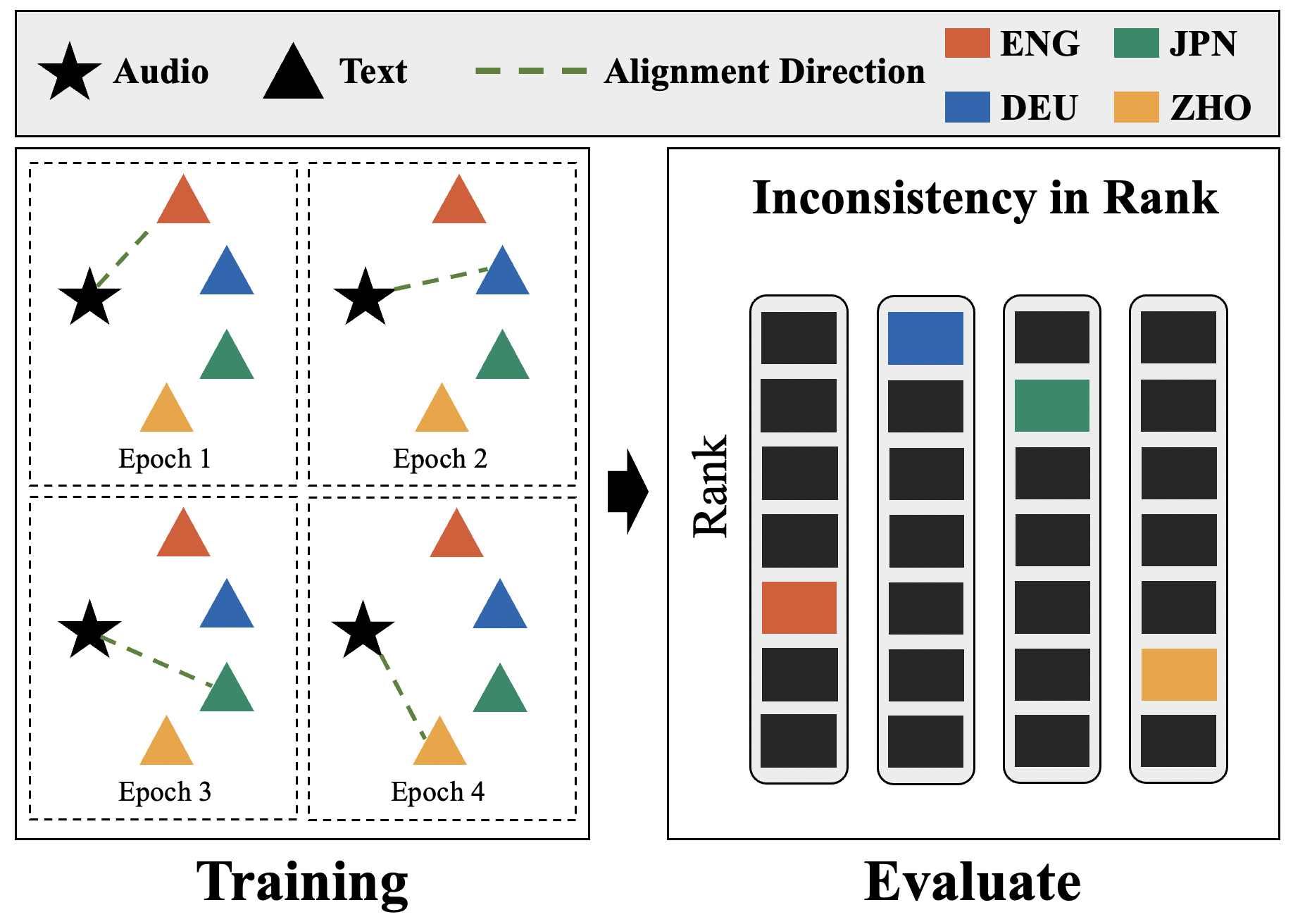}
    \caption{\textbf{An illustration of inconsistency issue in current ML-ATR scheme}.}
    \label{Fig:inconsistency illustration}
\end{figure}

To our knowledge, the existing mainstream ML-ATR scheme has a model training process as shown in Fig. \ref{Fig:inconsistency illustration}, which pairs audio with randomly selected linguistic text in each epoch. This may not allow the model to learn the embedding space of audio and multilingual texts very well, which reduces retrieval recall and makes it difficult to obtain the same retrieval results for audio and multilingual text instances in different languages.


In this paper, we theoretically analyze the causes of the inconsistency problem in ML-ATR. We first visualize the inconsistency problem in terms of the modal alignment direction error. The alignment direction error leads to the gradient error, which in turn invites the model weights to fail to converge to the optimal weights for multilingual modal alignment during the training process. We further heuristically derive theoretical upper bounds on the weight errors to quantify the adverse effects of inconsistency on the model weights. We analyze the composition of the weight error upper bound and conclude that the root cause of the error inconsistency is the data distribution error in training.


Based on the theoretical analysis, we propose a scheme to mitigate the inconsistency of ML-ATR, called ATRI. ATRI consists of two training strategies: 1-to-K Contrastive Learning (KCL) for the retrieval-performance-first scenario, and Audio-English Co-Anchor Contrastive Learning (CACL) for the overhead-first scenario. KCL theoretically eliminates the data distribution errors in each training epoch, thus achieving state-of-the-art performance in recall and consistency metrics. CACL aligns the other languages with audio and English text to correct the modal alignment direction and reduce the data distribution error. Compared to existing ML-ATR schemes, CACL improves retrieval recall and consistency while offering advantages in training time and GPU memory overhead over KCL.


Our contributions are shown below:
\begin{itemize}
    \item We analyze the inconsistency in terms of analyzing the modal alignment direction error and weighting error, and demonstrate an upper bound on the weighting error. We further conclude that the root cause of the inconsistency of existing ML-CLAP schemes lies in the distribution error of the training data.
    \item We propose ATRI, which solves the cross-lingual inconsistency problem in ML-ATR by reducing the data distribution error and correcting the modality alignment direction. ATRI contains the CACL and KCL training strategies for overhead-first and performance-first requirements, respectively.
    \item We evaluate the proposed scheme using the AudioCaps and Clotho datasets translated by Deepseek. The results show that ATRI effectively improves recall and consistency in both monolingual English ATR and ML-ATR tasks, achieving state-of-the-art performance.
\end{itemize}

\section{Related Work}
Audio-text retrieval (ATR) \cite{lou2022audio,xie2024gpa,xin2024diffatr} is a task that matches audio with text, which has seen significant advancements and widespread applications in recent years. The prevailing approach involves constructing a shared embedding space for audio and text, enabling seamless feature alignment and retrieving results based on similarity rankings. Widely adopted methods include CLIP-inspired \cite{yu2022coca,li2022blip} contrastive language-audio pertaining (CLAP) \cite{elizalde2023clap,wu2022wav2clip,guzhov2022audioclip}. Wu et al. \cite{wu2023large} introduce a feature fusion mechanism and a keyword-description enhancement strategy to enable the CLAP model to handle variable-length audio inputs and improve performance. Silva1 et al. \cite{silva2023collat} propose a framework that learns an audio understanding model by locking the language model parameters and employing an audio-text alignment pretraining objective for fine-grained audio comprehension. Ghosh et al. \cite{ghosh2023compa} design modular contrastive loss for difficult negative samples to enhance the fine-grained understanding of the CLAP model. In the next year, Ghosh et al. \cite{ghosh2025reclap} further enhance the model's understanding of real scene sounds by rewriting audio description text.

Existing ATR methods predominantly focus on English-centric monolingual tasks, with few solutions for multilingual scenarios \cite{yan2024bridging}. The scarcity of large-scale, accurately annotated non-English audio-caption datasets has led current ML-ATR methods to rely heavily on machine translation \cite{tiedemann2020opus,nllb2022no} to convert English datasets into multilingual versions. This translation-based strategy \cite{cousin2023multilingual,yan2024bridging} has demonstrated its effectiveness in enhancing datasets for multilingual use, significantly improving the recall performance of ATR systems. 

However, existing ML-ATR scheme \cite{yan2024bridging} uses audio-text pairs with randomly selected languages for training. As analyzed in Sect. \ref{Sect:Mathematical Demonstration about Inconsistency}, the training method employed presents significant challenges in achieving convergence to the optimal weights. This difficulty not only exacerbates issues related to inconsistent cross-lingual retrieval result, but also leads to a degradation in the retrieval performance, particularly in terms of both recall and accuracy.
\section{Definition and Inconsistency Analysis}
\label{Sect:Mathematical Demonstration about Inconsistency}
\subsection{Formal Definition of ML-ATR}
Audio-text retrieval is the task of learning cross-modality alignment between audio and multilingual text captions. Contrastive learning \cite{ru2023imbalanced,zhuang2025vargpt} has become the most effective method for learning expressive cross-modality embedding spaces.

Denote a dataset $D=\{(a_i, t_{i1},...t_{iK})\}_{i=1}^N$ as a multilingual audio text retrieval dataset, where $N$ denotes the size of dataset, $K$ refers the total language number in the dataset, $a_i$ denotes the audio in $i$-th data, $t_{ik}$ denotes the $k$-th language in $i$-th data. Given an audio encoder $f_\theta (\cdot)$ and a multilingual text encoder $g_\phi(\cdot)$, we denote the joint probability distribution as:

\begin{equation}
\label{Eq:origin distribution}
\small
    \begin{aligned}
        p(a_i,t_{ik})= \frac{\exp\left(s(f_\theta(a_i), g_\phi(t_{ik})) / \tau \right)}
  {\sum_{j=1}^N\sum_{l=1}^K \exp\left(s(f_\theta(a_j), g_\phi(t_{jl})) / \tau \right)},
    \end{aligned}
\end{equation}

\begin{equation}
\label{Eq:origin distribution}
\small
    \begin{aligned}
        p(a_i,t_{i})= \frac{\exp\left(s(f_\theta(a_i), g_\phi(t_{i})) / \tau \right)}
  {\sum_{j=1}^N \exp\left(s(f_\theta(a_j), g_\phi(t_{j})) / \tau \right)},
    \end{aligned}
\end{equation}

$s(\cdot)$ denotes the cosine similarity between audio and text embedding. The ideal optimization function of learning the embedding space is

\begin{equation}
\small
\begin{aligned}
\max_{\theta, \phi}\sum^{N}_{i=1}\sum^{K}_{k=1}p(a_i,t_{ik}) \mathbb{E}_{(a_i,t_{ik})}[log\ p(a_i,t_{ik})].
\end{aligned}
\end{equation}

However, instead of training all the languages of a piece of data in an epoch, the existing ML-ATR scheme randomly selects the text of a language to do the training. For each epoch $e$, a set of random numbers $Q=\{q_1,... .q_N\},q_i\stackrel{R}{\leftarrow}\{1,...K\}$. The optimization function they used is formalized as:

\begin{equation}
\label{Eq:error distribution}
\small
    \begin{aligned}
        p_e'(a_i,t_{iq_i})= \frac{\exp\left(s(f_\theta(a_i), g_\phi(t_{iq_i})) / \tau \right)}
  {\sum_{j=1}^N \exp\left(s(f_\theta(a_j), g_\phi(t_{jq_j})) / \tau \right)},
    \end{aligned}
\end{equation}

\begin{equation}
\small
\begin{aligned}
\max_{\theta, \phi}\sum^{N}_{i=1}p_e'(a_i,t_{iq_i}) \mathbb{E}_{(a_i,t_{iq_i})}[log\ p_e'(a_i,t_{iq_i})].
\end{aligned}
\end{equation}

The probability distribution $p_e'(a_i,t_{iq_i})$ of their scheme is not the same as the original probability distribution $p(a_i,t_{ik})$. This results in a model that does not fit the training data perfectly, making modality alignment ineffective, which in turn results in reduced recall and inconsistency problems.

\begin{figure}[htbp]
    \centering
    \includegraphics[scale=0.12]{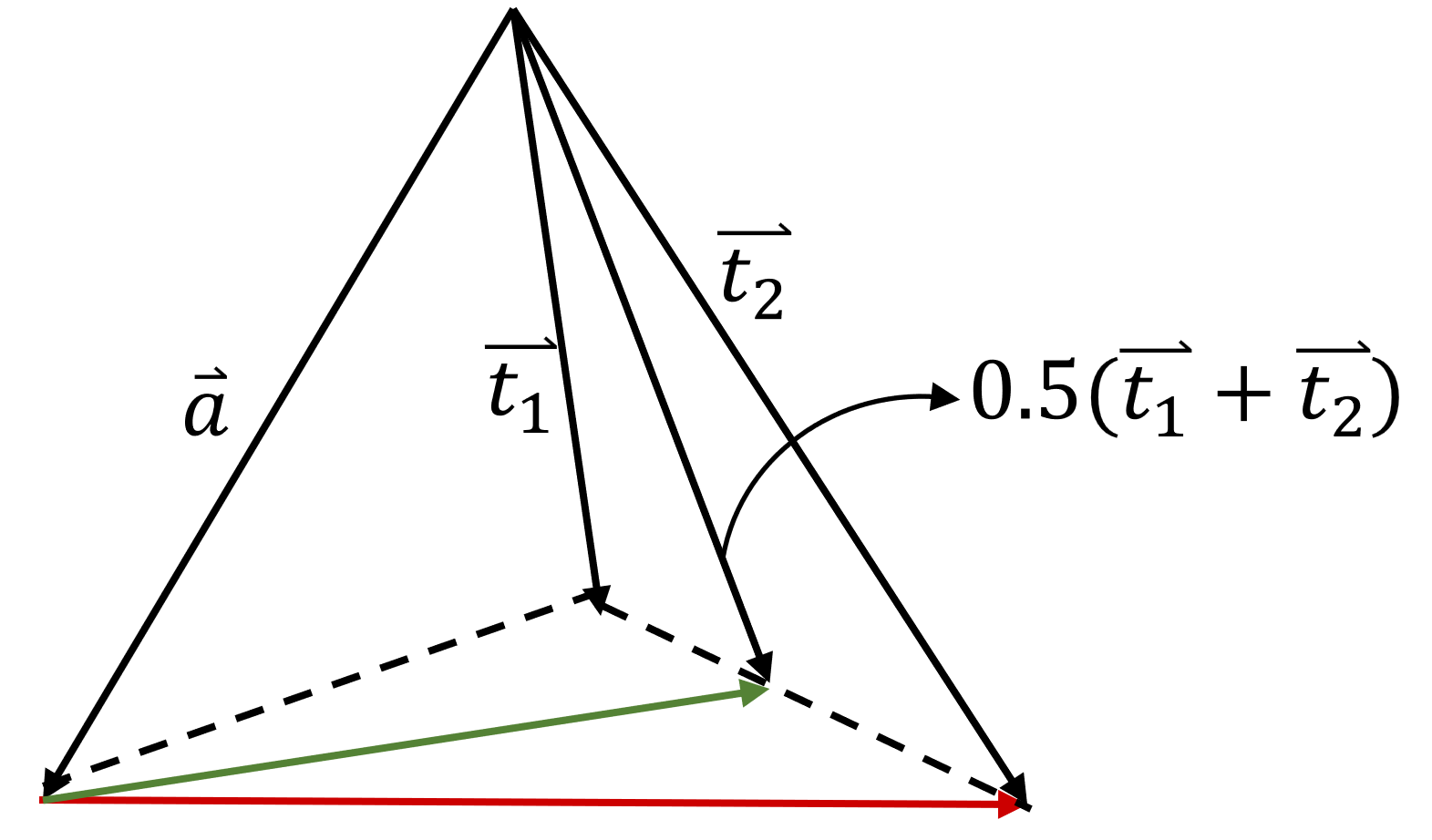}
    \caption{\textbf{A visual illustration of inconsistency due to modality alignment errors}.}
    \label{Fig:modality alignment error}
\end{figure}

\subsection{Analysis of the Inconsistency Issue}
\label{Sect:Description of the Inconsistency Issue}
We first analyze the issue of inconsistency from the perspective of modality alignment directional errors. As shown in Fig. \ref{Fig:modality alignment error}, an intuitive example of modality alignment error is illustrated. Consider a simple case of bilingual audio-text retrieval, let the embedding of an audio sample be $\vec{a}$, and the embeddings of the corresponding texts in two languages be $\vec{t_1}$ and $\vec{t_2}$. Ideally, the audio embedding $\vec{a}$ should be aligned with arithmetic mean embedding $\frac{1}{2}(\vec{t_1} + \vec{t_2})$ (indicated by the green arrow). We show why the arithmetic mean of multilingual embedding is the optimal alignment direction in Appendix \ref{Appe:Analysis Alignment Direction}. However, in existing ML-ATR schemes, the audio embedding is only aligned with the text embedding of a randomly selected language within each epoch. For instance, if the selected language is $2$, the audio embedding $\vec{a}$ will be aligned solely towards $\vec{t_2}$ (indicated by the red arrow). The angle between the red and green arrows is the modality alignment direction error, which makes the audio and multilingual text modes not well aligned.

Incorrect alignment introduces noise to the gradient, leading to errors between the model weights and their optimal values, making the model's retrieval recall and consistency metrics degrade. We give a theoretical weight error upper bound and analyze its composition to mitigate the inconsistency problem and improve retrieval recall. The detailed proof can be found in Appendix \ref{Appe:Proof of Weight Error Upper Bound}.


We assume that the optimization algorithm is stochastic gradient descent (SGD) \cite{ru2025we} to heuristically analyse the upper bound of the weight error. Given that the number of training steps per epoch $T$, the data distribution obtained by randomly sampling the language according to the existing ATR scheme is denoted as $p_e'$, and the original data distribution is denoted as $p$. $\mathbf w’_{eT}$ denotes the model weight in the $T$-th step under the $e$-th epoch trained with the data distribution $p'_e$, whereas $\mathbf w_{eT}$ denotes the weight that is trained with the data distribution $p$. If the gradient $\nabla_\mathbf w\mathbb{E}_{(a,t)}[log\ p(a,t)]$ is $\lambda_{(a,t)}$-Lipschitz \cite{bethune2023dp}, then we have the following inequality for weight error upper bound:
\begin{equation}
\label{Eq:weight error}
\small
\begin{aligned}
&||\mathbf w_{eT}-\mathbf w'_{eT}||\\
\leq & a^T||\mathbf w_{(e-1)T}-\mathbf w'_{(e-1)T}||+\\
&\eta \sum_{(a,t)}||p(a,t)-p'_e(a,t)||\sum^{T-1}_{j=1}(a^jg_{max}(\mathbf w_{eT-1-j})),
\end{aligned}
\end{equation}

\begin{equation}
\small
\begin{aligned}
g_{max}(\mathbf w)=max_{(a,t)}||\nabla_\mathbf w\mathbb{E}_{(a,t)}[log\ p(a,t)]||,
\end{aligned}
\end{equation}

\begin{equation}
\small
\begin{aligned}
a=1+\eta\sum_{(a,t)}p'_e(a,t)\lambda_{(x,y)}.
\end{aligned}
\end{equation}

\textbf{Note}: The weight $\mathbf w$ consists of the parameter $\theta$ for the audio encoder $f_\theta$ and the parameter $\phi$ for the multilingual text encoder $g_\phi$ in ML-ATR. The data distributions $p$ and $p'_e$ correspond to the Eq. \eqref{Eq:origin distribution} and \eqref{Eq:error distribution}, respectively. For simplicity, we denote $(a,t)$ as all audio-text pairs in the batch of the $T$-th step, where the text $t$ can be in any one of the languages. $\sum_{(a,t)}||p(a,t)-p'_e(a,t)||$ denotes the data distribution error in the batch at step $T$.

Detailed proof of Eq \eqref{Eq:weight error} can be found in Appendix \ref{Appe:Proof of Weight Error Upper Bound}. Based on Eq. \eqref{Eq:weight error}, we have the following results:

\begin{itemize}
    \item Intuitively, the weight error $||\mathbf w_{eT}-\mathbf w'_{eT}||$ comes from two main sources. One is the weight error after the $(e-1)$-th epoch, i.e. $||\mathbf w'_{(e-1)T}-\mathbf w_{(e-1)T}||$. The other is caused by the probabilistic distances of the data distributions, i.e. $\sum_{(a,t)}||p'_e(a,t)-p(a,t)||$. Since $a\geq 1$, the error from both sources increases with epoch and step. In addition, the weight error is also affected by the learning rate $\eta$, the number of training steps $T$ and the maximum gradient $g_{max}(\mathbf w_{eT-1-j})$.
    \item Further expansion of Eq. \eqref{Eq:weight error} shows that the weighting error arises from the data distribution error of each epoch. Expanding $||\mathbf w_{(e-1)T}-\mathbf w'_{(e-1)T}||$ in Eq. \eqref{Eq:weight error}, we find it consist of $||\mathbf w_{(e-2)T}-\mathbf w'_{(e-2)T}||$ and $||p(a,t)-p'_{e-1}(a,t)||$. Further expanding Eq. \eqref{Eq:weight error} to the weight error in $1$-th epoch, it can be concluded that the weight error of the existing ML-ATR scheme comes from the data distribution error $\sum^e_{i=1}\sum_{(a,t)}||p(a,t)-p'_i(a,t)||$ due to the randomly selected languages in each epoch. We can mitigate the inconsistency problem and improve the recall by reducing the weight error upper bound by reducing the data distribution error for each epoch.
\end{itemize}
\section{Proposed ML-ATR Scheme}
We propose two methods to reduce the data distribution error during training. One is 1-to-K contrastive learning, which has a higher memory overhead. The other is audio-English co-anchor contrastive learning, which achieves performance close to 1-to-K Contrastive Learning while approximating the memory overhead to the existing ML-ATR scheme. Here are the details of the two methods.

\subsection{1-to-K Contrastive Learning}
Building on our theoretical analyses, we conclude that reducing data distribution error is critical for addressing the cross-lingual inconsistency problem in multilingual audio-text retrieval. To achieve this, we propose 1-to-K Contrastive Learning (KCL), a training strategy that replaces random language sampling with the simultaneous use of all $K$ linguistic texts corresponding to each audio instance. This approach theoretically eliminates data distribution error, corrects modal alignment direction, and significantly enhances both the recall and consistency of retrieval performance. The loss function $\mathcal{L}_{kcl}$ for the proposed 1-to-K Contrastive Learning in ML-ATR is defined as follows:

\begin{equation}
\small
\begin{aligned}
\mathcal{L}_{kcl}=\frac{1}{2NK}(\mathcal{L}^{a2t}_{kcl}+\mathcal{L}^{t2a}_{kcl}).
\end{aligned}
\end{equation}

The loss function $\mathcal{L}^{at}_{kcl}$ consists of two parts, $\mathcal{L}^{a2t}_{kcl}$ and $\mathcal{L}^{t2a}_{kcl}$, and they are calculated as follows:

\begin{equation}
\small
\begin{aligned}
\mathcal{L}^{a2t}_{kcl}=-\sum^K_{k=1}\sum^N_{i=1}log\frac{\exp(s(f_\theta(a_i),g_\phi(t_{ik}))/\tau)}{\sum^N_{j=1}\exp(s(f_\theta(a_i),g_\phi(t_{jk}))/\tau)},
\end{aligned}
\end{equation}

$\mathcal{L}^{a2t}_{kcl}$ denotes the contrastive learning loss function from audio to multilingual text.

\begin{equation}
\small
\begin{aligned}
\mathcal{L}^{t2a}_{kcl}=-\sum^K_{k=1}\sum^N_{i=1}log\frac{\exp(s(g_\phi(t_{ik}),f_\theta(a_i))/\tau)}{\sum^N_{j=1}\exp(s(g_\phi(t_{ik}),f_\theta(a_j))/\tau)},
\end{aligned}
\end{equation}

$\mathcal{L}^{t2a}_{kcl}$ denotes the contrastive learning loss function from multilingual text to audio.

$K$ is the number of languages and $N$ is the number of data instances. As shown in Tab. \ref{Tab:overhead}, including multiple multilingual texts in 1-to-K contrastive learning increases GPU memory usage and training time. In practical ML-ATR applications, supporting more languages amplifies these overheads compared to existing schemes.

To address this, we further propose CACL, which improves retrieval consistency and recall without significantly increasing overhead.

\subsection{Audio-English Co-Anchor Contrastive Learning}
To reduce the weighting error with as little increase in training time and GPU memory consumption as possible, we propose audio-English co-anchor contrastive learning (CACL). During the training process, each data takes its audio, English text, and text in other random languages and does contrastive learning with each other. 

For each epoch, given a set of random numbers $Q=\{q_1,...q_N\},q_i\stackrel{R}{\leftarrow}\{2,...K\}$, get the triplet of the training data $(a_i,t_{i1},t_{iq_i})$, where $a_i$ denotes $i$-th audio, $t_{i1}$ denotes the English text, and $t_{iq_i}$ denotes the text of $q_i$-th language. We have the training loss $\mathcal{L}_{cacl}$ shown below:

\begin{equation}
\small
\begin{aligned}
\mathcal{L}_{cacl}=\frac{1}{6N}(\mathcal{L}^{ae}_{cacl}+\mathcal{L}^{at}_{cacl}+\mathcal{L}^{et}_{cacl}).
\end{aligned}
\end{equation}

The loss function $\mathcal{L}_{cacl}$ consists of three components $\mathcal{L}^{ae}_{cacl},\mathcal{L}^{at}_{cacl},\mathcal{L}^{et}_{cacl}$. All three components are based on the following general contrastive learning loss formulation:

\begin{equation}
\small
\begin{aligned}
\mathcal{L}^{uv}_{cacl}=&-\sum^N_{i=1}log\frac{\exp(s(u_i,v_i)/\tau)}{\sum^N_{j=1}\exp(s(u_i,v_j)/\tau)}\\
&-\sum^N_{i=1}log\frac{\exp(s(v_i,u_i)/\tau)}{\sum^N_{j=1}\exp(s(v_i,u_j)/\tau)},
\end{aligned}
\end{equation}
where $u_i$ and $v_i$ represent input embeddings from different modalities or languages. The three components are defined as follows:

\begin{itemize}
    \item \textbf{Audio-English Alignment} ($\mathcal{L}^{ae}_{cacl}$): 
    
    $u_i=f_\theta(a_i)$ represents audio embeddings, and $v_i=g_\phi(t_{i1})$ represents English text embeddings.
    \item \textbf{Audio-Multilingual Alignment} ($\mathcal{L}^{at}_{cacl}$): $u_i=f_\theta(a_i)$ represents audio embeddings, and $v_i=g_\phi(t_{iq_i})$ represents text embeddings in a randomly selected language.
    \item \textbf{English-Multilingual Alignment} ($\mathcal{L}^{et}_{cacl}$): $u_i=g_\phi(t_{i1})$ represents English text embeddings, and $v_i=g_\phi(t_{iq_i})$ represents text embeddings in a randomly selected language.
\end{itemize}







The effectiveness of audio-English CACL can be explained from two perspectives:
\begin{itemize}
    \item From the perspective of modality alignment (Fig. \ref{Fig:modality alignment error}), the loss function $\mathcal{L}^{et}_{cacl}$ in CACL brings embeddings of English and other languages closer, reducing the distance between the text embedding $\vec{t_1},\vec{t_2}$ and the mean $\frac{1}{2}(\vec{t_1}+\vec{t_2})$ and minimizing the deviation in the modality alignment direction of audio and text.
    \item From the perspective of data distribution error $\sum_{(a,t)}||p(a,t)-p'_e(a,t)||$ in Eq. \eqref{Eq:weight error}, CACL's loss functions $\mathcal L^{ae}_{cacl}, \mathcal L^{at}_{cacl}$ ensures that the model learns more pairs of audio texts in an epoch. The text in them also contains a large percentage of high-quality English text. It makes the data distribution in CACL closer to the original one, and reduces the weight error of the model.
\end{itemize}

Note that in CACL, the number of texts used for training in each epoch does not increase with the number of languages, which effectively reduces both GPU memory and time overhead in ML-ATR scenarios with a large number of languages. Our experimental results illustrate that CACL approximates the training time and explicit memory overhead of existing ML-ATR schemes, yet achieves recall and consistency metrics close to those of 1-to-K comparative learning.
\section{Experiments}
\subsection{Dataset}
We employ the AudioCaps \cite{kim2019audiocaps}, and Clotho \cite{drossos2020clotho} for our experiments. AudioCaps includes around 49,000 audio samples, each lasting about 10 seconds. Each audio is paired with a single sentence in the training set, while in both the validation and test sets, each audio has five associated sentences. The Clotho dataset consists of 6,974 audio samples, each ranging from 15 to 30 seconds long and annotated with five sentences. It is split into 3,839 training samples, 1,045 validation samples, and 1,045 test samples. 

Additionally, to assess our scheme's performance in the ML-ATR task, we use the Deepseek \cite{bi2024deepseek} API to translate the text from AudioCaps and Clotho into seven widely spoken languages, including French (fra), German (deu), Spanish (spa), Dutch (nld), Catalan (cat), Japanese (jpn), and Chinese (zho). We evaluate the degree of disillusion in the test set in Appendix \ref{Appe:Multilingual Test Set Hallucination Assessment}.

\subsection{Models}
\textbf{Audio Encoder}: 
We utilize the recently proposed CED-Base model \cite{dinkel2024ced}, a vision transformer with 86 million parameters for the Audio Encoder. Trained on Audioset through knowledge distillation from a large teacher ensemble, the model processes 64-dimensional Mel-spectrograms derived from a 16 kHz signal. It then extracts non-overlapping 16 × 16 patches from the spectrogram, resulting in 248 patches over a 10-second input (4 × 62).
\\
\textbf{Text Encoder}:
The key to multilingual audio-text retrieval is the text encoder's ability to handle texts in multiple languages. In this work, we focus solely on the SONAR-TE model \cite{duquenne2023sonar}. SONAR-TE model significantly outperforms the LaBSE model \cite{feng2020language} under the multilingual xsim and xsim++ tasks, and can extract sentence embeddings more accurately. The fixed-size text representation is derived by pooling the token-level outputs from the encoder. In the following sections, SONAR refers specifically to the text encoder.

\subsection{Setup}
We use ML-CLAP \cite{yan2024bridging} as the baseline, which is state-of-the-art for ML-ATR tasks. To have a fair comparison, the model is initialized using the pre-trained weights of ML-CLAP and is further fine-tuned on our multilingual AudioCaps and Clotho datasets using three training methods: ML-CLAP, proposed CACL, and proposed KCL.

All models were fine-tuned for 10 epochs on a single A100 80GB PCIe GPU with a batch size of 24, a learning rate of $5 \times 10^{-6}$, using the Adam optimizer. The temperature hyperparameter $\tau$ was set to 0.07 for all configurations. The audio was sampled at $1.6\times 10^{4}$. We selected the model with the best recall performance during the fine-tuning period for each scheme to perform the experiments.

\subsection{Evaluation Metric}
We use the recall of rank k (R@k) and the average precision of rank 10 (mAP10) as the metrics for the retrieval performance of the model to show that reducing data distribution errors improves the retrieval performance in each language. R@k refers to the fact that for a query, R@k is 1 if the target-value item occurs in the first k retrieved items, and 0 otherwise. mAP10 calculates the average precision of all the queries among the first 10 retrieved results. With these two metrics, we can comprehensively evaluate the retrieval performance of the model on multilingual datasets. 

To assess the consistency of the embedding space across languages, we use three metrics: embedding space gap $\vec{\triangle}_{gap,k}$ \cite{liang2022mind}, average embedding distance $\vec{\triangle}_{dis,k}$, mean rank variance (MRV). The computation of $\vec{\triangle}_{gap,k}$, $\vec{\triangle}_{dis,k}$ and MRV is shown below:

\begin{equation}
\small
    \begin{aligned}
        \vec{\triangle}_{gap,k}=\frac 1 N \sum^N_{i=1}g_\phi(t_{i1})-\frac 1 N \sum^N_{i=1}g_\phi(t_{ik}),
    \end{aligned}
\end{equation}

\begin{equation}
\small
    \begin{aligned}
        \vec{\triangle}_{dis,k}=\frac 1 N \sum^N_{i=1}||g_\phi(t_{i1})-g_\phi(t_{ik})||,
    \end{aligned}
\end{equation}

\begin{equation}
\small
    \begin{aligned}
        MRV=\frac 1 {NK} \sum^N_{i=1}\sum^K_{k=1}|Rank_{ik}-\overline{Rank_j}|^2.
    \end{aligned}
\end{equation}

$\vec{\triangle}_{gap,k}$ and $\vec{\triangle}_{dis,k}$ denotes the embedding space gap and average embedding distance between English and $k$-th language respectively. $Rank_{ik}$ denotes the similarity ranking of the $k$-th language under the $i$-th data, and $\overline{Rank_i}$ denotes the average similarity ranking under the $i$-th data.

\begin{table*}[ht]
\caption{Recall and precision results for baseline and our method under multilingual AudioCaps and Clotho dataset}
\small
\centering
\begin{tabular}{c|c|ccc|ccc|ccc|ccc}
\hline
\multirow{3}{*}{\rotatebox{90}{\textbf{Scheme}}} & \multirow{3}{*}{\textbf{Lang}} & \multicolumn{6}{c|}{\textbf{AudioCaps}} & \multicolumn{6}{c}{\textbf{Clotho}}\\ 
\cline{3-14} & & \multicolumn{3}{c|}{T2A} & \multicolumn{3}{c|}{A2T} & \multicolumn{3}{c|}{T2A} & \multicolumn{3}{c}{A2T}\\
\cline{3-14}
 & & R@1 & R@5 & mAP10 & R@1 & R@5 & mAP10 & R@1 & R@5 & mAP10 & R@1 & R@5 & mAP10 \\ \cline{1-14}
\multirow{9}{*}{\rotatebox{90}{ML-CLAP}} & eng & 47.31 & 80.65 & 61.44 & 64.91 &	90.54 &	38.62 &	25.98 &	54.5 & 38.15 & 34.03 &	61.05 &	21.19 \\ 
& fra & 45.88 &	78.92 &	60.01 &	61.65 &	89.39 &	37.90 &	24.42 &	52.51 &	36.24 &	30.95 &	57.59 &	19.66 \\ 
& deu & 45.60 &	79.49 &	59.93 &	62.65 &	88.76 &	37.88 &	24.08 &	52.61 &	36.40 &	31.62 &	57.40 &	19.39 \\ 
& spa & 45.00 &	79.32 &	59.62 &	63.04 &	88.86 &	37.38 &	24.05 &	52.75 &	36.22 &	31.43 &	57.98 &	19.65 \\ 
& nld & 45.88 &	79.64 &	59.92 &	62.50 &	90.33 &	37.72 &	23.88 &	51.53 &	35.73 &	31.40 &	57.98 &	19.58 \\ 
& cat & 44.36 &	77.89 &	58.58 &	61.65 &	87.60 &	36.43 &	22.83 &	50.84 &	34.80 &	30.91 &	56.43 &	18.26 \\ 
& jpn & 43.04 &	76.86 &	57.54 &	59.45 &	87.81 &	35.20 &	23.04 &	50.34 &	34.89 &	31.43 &	56.55 &	18.77 \\ 
& zho & 41.70 &	74.72 &	55.74 &	53.67 &	84.76 &	33.38 &	21.65 &	48.84 &	33.53 &	28.41 &	56.14 &	17.26 \\ \cline{2-14}
& avg & 44.84 &	78.43 &	59.09 &	61.19 &	88.50 &	36.81 &	23.84 &	51.74 &	35.74 &	31.27 &	57.64 &	19.22 \\ \hline

\multirow{9}{*}{\rotatebox{90}{our CACL}} & eng & 49.05 &	82.14 &	63.07 &	66.31 &	\textbf{91.49} &	39.41 &	26.36 &	55.19 &	38.68 &	34.71 &	61.34 &	\textbf{21.57} \\ 
& fra & 46.86 &	79.97 &	60.83 &	63.23 &	89.48 &	37.92 &	\textbf{24.90} &	\textbf{53.09} &	36.67 &	\textbf{32.40} &	58.55 &	19.85 \\ 
& deu & 46.21 &	80.08 &	60.62 &	63.13 &	\textbf{89.91} &	38.14 &	24.51 &	52.86 &	36.52 &	\textbf{33.36} &	58.07 &	19.49 \\ 
& spa & 46.68 &	80.52 &	60.90 &	63.23 &	\textbf{90.12} &	37.45 &	\textbf{24.59} &	52.71 &	\textbf{36.72} &	32.40 &	58.17 &	19.75 \\ 
& nld & 47.41 &	80.23 &	61.22 &	63.23 &	\textbf{90.86} &	37.95 &	24.15 &	51.75 &	36.05 &	32.21 &	58.65 &	19.5 \\ 
& cat & 45.27 &	78.61 &	59.43 &	61.23 &	88.44 &	36.49 &	23.28 &	51.42 &	35.17 &	30.67 &	56.05 &	18.67 \\ 
& jpn & 44.76 &	78.50 &	58.97 &	61.55 &	88.67 &	34.91 &	23.36 &	51.53 &	35.28 &	\textbf{31.82} &	\textbf{58.26} &	18.99 \\ 
& zho & 42.01 &	76.02 &	56.23 &	56.40 &	86.65 &	33.93 &	22.50 &	49.42 &	34.01 &	27.69 &	\textbf{57.59} &	17.48 \\ \cline{2-14}
& avg & 46.03 &	79.50 &	60.15 &	62.28 &	89.45 &	37.02 &	24.20 &	52.24 &	36.27 &	31.90 &	58.33 &	19.41
 \\ \hline

\multirow{9}{*}{\rotatebox{90}{our KCL}} & eng & \textbf{49.68} &	\textbf{82.44} &	\textbf{63.34} &	\textbf{66.59} &	91.34 &	\textbf{40.52} &	\textbf{26.67} &	\textbf{55.46} &	\textbf{38.97} &	\textbf{36.34} &	\textbf{64.13} &	21.36 \\
& fra & \textbf{47.79} &	\textbf{80.52} &	\textbf{61.53} &	\textbf{63.41} &	\textbf{}\textbf{89.57} &	\textbf{39.21} &	24.61 &	52.73 &	\textbf{36.79} &	31.82 &	\textbf{60.76} &	\textbf{20.02} \\ 
& deu & \textbf{47.81} &	\textbf{80.81} &	\textbf{61.78} &	\textbf{63.34} &	89.28 &	\textbf{39.02} &	\textbf{24.90} &	\textbf{53.25} &	\textbf{37.02} &	33.17 &	\textbf{59.61} &	\textbf{19.90} \\ 
& spa & \textbf{47.33} &	\textbf{80.67} &	\textbf{61.49} &	\textbf{63.76} &	89.39 &	\textbf{38.73} &	24.31 &	\textbf{52.96} &	36.55 &	\textbf{33.36} &	\textbf{61.25} &	\textbf{20.27} \\ 
& nld & \textbf{47.92} &	\textbf{}\textbf{80.76} &	\textbf{61.70} &	\textbf{63.55} &	90.52 &	\textbf{39.14} &	\textbf{24.53} &	\textbf{52.51} &	\textbf{36.61} &	\textbf{33.55} &	\textbf{62.30} &	\textbf{19.98} \\ 
& cat & \textbf{46.44} &	\textbf{79.62} &	\textbf{60.42} &	\textbf{62.71} &	\textbf{89.49} &	\textbf{37.65} &	\textbf{23.67} &	\textbf{51.86} &	\textbf{35.70} &	\textbf{31.53} &	\textbf{57.98} &	\textbf{1}\textbf{8.90} \\ 
& jpn & \textbf{45.27} &	\textbf{78.86} &	\textbf{59.49} &	\textbf{62.28} &	\textbf{89.16} &	\textbf{36.81} &	\textbf{23.65} &	\textbf{52.17} &	\textbf{35.68} &	31.25 &	57.50 &	\textbf{19.49} \\ 
& zho & \textbf{42.25} &	\textbf{76.38} &	\textbf{56.75} &	\textbf{57.66} &	\textbf{87.28} &	\textbf{34.79} &	\textbf{23.09} &	\textbf{49.90} &	\textbf{34.60} &	\textbf{30.48} &	56.34 &	\textbf{17.85} \\ \cline{2-14}
& avg & \textbf{46.81} &	\textbf{80.00} &	\textbf{60.81} &	\textbf{62.91} &	\textbf{89.50} &	\textbf{38.23} &	\textbf{24.42} &	\textbf{52.60} &	\textbf{36.49} &	\textbf{32.68} &	\textbf{59.98} &	\textbf{19.72} \\ \hline
\end{tabular}
\label{Tab:recall and meanr}
\end{table*}

\subsection{Evaluation Result of Recall and Precision}
We present a detailed numerical comparison analysis of the experiment results in Tab \ref{Tab:recall and meanr}, focusing on the performance improvements of our proposed methods, CACL and KCL, over the baseline ML-CLAP across various languages and datasets. We supplement evaluation with the LaBSE model as a text encoder in Appendix \ref{Appe:Supplemental Evaluation with LaBSE Text Encoder}.

\subsubsection{Analysis of Evaluation Results}
Overall, the proposed CACL and KCL consistently outperform ML-CLAP across the majority of languages and datasets in terms of recall at 1 (R@1), recall at 5 (R@5), and mean average precision at the top 10 results (mAP10) for both Text-to-Audio (T2A) and Audio-to-Text (A2T) tasks. Notably, our proposed KCL achieves state-of-the-art performance, delivering a 5\% improvement in R@1 for the English-oriented monolingual ATR task and a 4.3\% improvement in R@1 for the multilingual ATR task compared to ML-CLAP. This experimental result corroborates our theoretical analysis of the weighting error in Sect. \ref{Sect:Mathematical Demonstration about Inconsistency}. Here is the detailed analysis:

CACL's average metrics across languages are higher than ML-CLAP, while KCL's average metrics across languages have further improvement compared to CACL. Our theoretical analyses in Sect can explain this phenomenon. \ref{Sect:Mathematical Demonstration about Inconsistency}:
\begin{itemize}
    \item CACL uses audio and text together as the anchor point for modality alignment in other languages, which can effectively reduce the data distribution error and modality alignment error, thus achieving better modality alignment results and improved metrics compared to ML-CLAP.
    \item Compared to CACL, which mitigates data distribution errors, KCL theoretically eliminates these errors. As a result, KCL achieves superior modality alignment compared to CACL, leading to further improvements in both recall and precision.
\end{itemize}

\subsubsection{Analysis of Special Situations}

\textbf{Occasional Metric Anomalies}: We observed occasional anomalies where a small proportion of KCL metrics were lower than CACL metrics, and some CACL metrics were lower than ML-CLAP metrics. We attribute these discrepancies to noise in the dataset. Specifically, the weight error in Eq. \eqref{Eq:weight error} represents the difference between the current and optimal model weights for fitting the training data. If the dataset is too noisy, the optimal weights may not improve the test set's performance. As a result, KCL and CACL, which have lower weight errors, may still underperform ML-CLAP on certain metrics. The higher frequency of such anomalies in the noisier Clotho dataset, compared to Audiocaps, supports this explanation. Given that these anomalies are rare among the results in Tab. \ref{Tab:recall and meanr}, we consider them acceptable and conclude that they do not impact the overall performance advantage of CACL and KCL in the ML-ATR task.

\textbf{Performance Gaps Across Languages}: The lower metrics for Japanese and Chinese in Tab. \ref{Tab:recall and meanr} are mainly due to their significant syntactic differences from other languages, making them harder for the model to learn. Expanding the dataset for these languages could improve the model's performance by providing more representative data.

\textbf{Better Replicated Performance}: Compared to the original paper, our replicated ML-CLAP model achieves significant improvements across all metrics, mainly due to differences in data quality. Compared to the text translated by the SONAR model \cite{duquenne2023sonar} used by baseline, the multilingual text we translated with LLM is of higher quality, which in turn can improve the retrieval performance of the model.

\begin{table}[ht]
\caption{Results of spatial differences in the embedding of other languages and English}
\small
\centering
\begin{tabular}{c|c|cc|cc}
\hline
\multirow{3}{*}{\rotatebox{90}{\textbf{Scheme}}} & \multirow{3}{*}{\textbf{Lang}} & \multicolumn{2}{c|}{\textbf{AudioCaps}} & \multicolumn{2}{c}{\textbf{Clotho}}\\ 
\cline{3-6} & & \multicolumn{2}{c|}{E2T} & \multicolumn{2}{c}{E2T}\\
\cline{3-6}
 & & Gap & Dis & Gap & Dis \\ \cline{1-6}
\multirow{8}{*}{\rotatebox{90}{ML-CLAP}} & fra & 0.199 & 0.094 & 0.120 & 0.301\\ 
& deu & 0.210 & 0.370 & 0.124 & 0.289\\ 
& spa & 0.147 & 0.290 & 0.117 & 0.284\\ 
& nld & 0.204 & 0.346 & 0.121 & 0.274\\ 
& cat & 0.151 & 0.357 & 0.121 & 0.307\\ 
& jpn & 0.237 & 0.445 & 0.123 & 0.353\\ 
& zho & 0.181 & 0.414 & 0.177 & 0.323\\ \cline{2-6}
& avg & 0.189 & 0.330 & 0.129 & 0.304\\ \hline

\multirow{8}{*}{\rotatebox{90}{our CACL}} & fra & 0.160 &  0.281 & 0.112 & 0.288\\ 
& deu & 0.194 & 0.334 & 0.103 & 0.261\\ 
& spa & 0.090 & 0.210 & 0.099 & 0.265\\ 
& nld & 0.172 & 0.325 & 0.106 & 0.255\\ 
& cat & 0.104 & 0.252 & 0.108 & 0.280\\ 
& jpn & 0.217 & 0.402 & 0.122 & 0.359\\ 
& zho & 0.192 & 0.381 & 0.159 & 0.352\\ \cline{2-6}
& avg & 0.161 & 0.312 & 0.115 & 0.294\\ \hline

\multirow{8}{*}{\rotatebox{90}{our KCL}} & fra & \textbf{0.145} & \textbf{0.274} & \textbf{0.094} & \textbf{0.261}\\ 
& deu & \textbf{0.155} & \textbf{0.290} & \textbf{0.084} & \textbf{0.231}\\ 
& spa & \textbf{0.081} & \textbf{0.192} & \textbf{0.084} & \textbf{0.230}\\ 
& nld & \textbf{0.148} & \textbf{0.285} & \textbf{0.072} & \textbf{0.204}\\ 
& cat & \textbf{0.092} & \textbf{0.245} & \textbf{0.087} & \textbf{0.243}\\ 
& jpn & \textbf{0.188} & \textbf{0.356} & \textbf{0.106} & \textbf{0.310}\\ 
& zho & \textbf{0.181} & \textbf{0.379} & \textbf{0.123} & \textbf{0.312}\\ \cline{2-6}
& avg & \textbf{0.141} & \textbf{0.288} & \textbf{0.092} & \textbf{0.255}\\ \hline
\end{tabular}
\label{Tab:embeddings gap}
\end{table}

\subsection{Evaluation Result of Consistency}
\subsubsection{Analysis of Embedding Space Consistency}
The results of the consistency metrics embedding space gap $\vec{\triangle}_{gap,k}$ and average embedding distance $\vec{\triangle}_{dis,k}$ are shown in Tab. \ref{Tab:embeddings gap}. In addition, we give a visualization of the embedding space in Appendix \ref{Appe:Embedding Space} and case analysis in Appendix \ref{Appe:Case Analysis} to further illustrate the effectiveness of ATRI in solving the consistency problem.

Smaller values of $\vec{\triangle}_{gap,k}$ and $\vec{\triangle}_{dis,k}$ indicate better alignment of a language's embedding space with English, leading to more consistent retrieval in the ML-ATR task. Compared to the baseline ML-CLAP, CACL achieves an average reduction of 12.9\% in Gap and 4.4\% in Dis, while KCL reduces Gap by 19.1\% and Dis by 14.3\%, demonstrating improved cross-language retrieval consistency.

\begin{table}[ht]
\caption{Results of Mean Rank Variance}
\small
\centering
\begin{tabular}{c|c|c}
\hline
\multirow{2}{*}{\textbf{Scheme}} & \textbf{AudioCaps} & \textbf{Clotho}\\ \cline{2-3}
 & MRV & MRV \\ \hline
ML-CLAP & 10.38 & 347.34 \\ \hline
CACL & 8.71 & 274.87 \\ \hline
KCL & \textbf{7.52} & \textbf{263.15} \\ \hline
\end{tabular}
\label{Tab:MRV}
\end{table}

\subsubsection{Analysis of Rank Consistency}
MRV quantifies the consistency of search rankings across languages, with lower values indicating more consistent results across languages. Unlike metrics based on embedding space, MRV offers a more direct assessment of model consistency in the ML-ATR task. As shown in Tab. \ref{Tab:MRV}, KCL achieves the lowest MRV, representing a 25.9\% reduction compared to ML-CLAP, while CACL achieves a 22.3\% reduction. This effectively shows that the inconsistency issue can be effectively mitigated by reducing the data distribution error.

We note that the MRV metrics under the Audiocaps dataset are significantly lower than Clotho's. This is due to the fact that the Clotho dataset is much noisier and more difficult to get consistent retrieval results across languages.

\begin{table}[ht]
\caption{Evaluation results in GPU memory overheads and time overheads}
\small
\centering
\begin{tabular}{c|cc|cc}
\hline
\multirow{2}{*}{\textbf{Scheme}} & \multicolumn{2}{c|}{\textbf{AudioCaps}} & \multicolumn{2}{c}{\textbf{Clotho}} \\ \cline{2-5}
 & GMO(MB) & TO(s) & GMO(MB) & TO(s)\\ \hline
ML-CLAP & 22172 & 3349 & 30912 & 1592\\ \hline
our CACL & 26788 & 3745 & 31528 & 1714\\ \hline
our KCL & 68256 & 4209 & 79480 & 1884\\ \hline
\end{tabular}
\label{Tab:overhead}
\end{table}

\subsection{Evaluation Results about Training Overhead}
Tab.\ref{Tab:overhead} summarises the GPU memory overhead (GMO) and time overhead (TO) during training for three scenarios: ML-CLAP, CACL, and KCL. KCL training requires simultaneous input of text in eight languages, which significantly increases overhead, resulting in a higher GMO of about 2.8 times and a 27\% increase in TO compared to ML-CLAP. In contrast, CACL inputs just twice as much text as ML-CLAP, resulting in a modest increase of about 10\% in both overheads. This makes CACL more suitable for scenarios that prioritize lower training overheads, while KCL is more suitable for applications that emphasize retrieval performance.
\section{Conlusion}
In this paper, we address the cross-lingual inconsistencies in ranking results observed in existing ML-ATR schemes. Through an analysis of modality alignment errors and weighting errors, we identify data distribution errors during training as a key factor impacting cross-lingual modality alignment, ultimately leading to retrieval inconsistencies. To address this, we propose two training strategies: KCL and CACL, designed for scenarios prioritizing retrieval performance and training overhead, respectively. Experimental results demonstrate that both CACL and KCL effectively enhance retrieval performance and consistency in ML-ATR tasks. Notably, KCL achieves state-of-the-art results in both English-oriented monolingual ATR and ML-ATR tasks. Furthermore, the proposed approach of mitigating data distribution errors to reduce inconsistencies holds potential for broader applications, including multilingual modality alignment in image and video modalities.

\section*{Limitation}
We acknowledge that the upper bound on the weighting error in Eq. \eqref{Eq:weight error} is heuristically proven for the SGD optimizer. For more complex optimizers such as Adam, giving a direct upper bound on the weighting error is difficult. But we provide proof of momentum error upper bound for Adam in the Appendix \ref{Appe:Migrating}, and show that our idea of reducing the data distribution error is still feasible under the Adam optimizer by showing that momentum error leads to weight error.

\section*{Acknowlegdements}
This work is supported by the Guangdong Provincial Key Laboratory of Ultra High Definition Immersive Media Technology(Grant No. 2024B1212010006).

\bibliographystyle{acl_natbib}
\bibliography{reference}

\appendix
\section{Multilingual Test Set Hallucination Assessment}
\label{Appe:Multilingual Test Set Hallucination Assessment}
Since we conducted our experiments on a multilingual audio text retrieval dataset translated with a large language model, to further illustrate the reliability of the experimental results, we evaluated the level of illusions in the test set. We use Deepseek V3 to re-translate the translated texts in the test set into English. Then use the Roberta-large model to evaluate the embedding average cosine similarity between the re-translated texts and the original English texts. The detailed experimental results are shown in \ref{tab:language_comparison}. It can be seen that the average similarity between each language and English reaches above 90\%, indicating that the semantics of the translated text and the original English text are very close, and that the illusions in the test set are sufficiently small.

\begin{table}[h]
\centering
\caption{Semantic Similarity between Re-translated Text and Original English Text on Test Set}
\begin{tabular}{lcc}
\hline
Language & AudioCaps & Clotho \\
\hline
French & 94.5 & 93.5 \\
Dutch & 95.0 & 94.2 \\
Spanish & 94.8 & 94.0 \\
German & 95.6 & 94.2 \\
Catalan & 91.1 & 92.2 \\
Japanese & 91.7 & 90.9 \\
Chinese & 90.7 & 90.3 \\
\hline
\end{tabular}
\label{tab:language_comparison}
\end{table}

\section{Supplemental Evaluation with LaBSE Text Encoder}
\label{Appe:Supplemental Evaluation with LaBSE Text Encoder}
We conduct complementary experiments using LaBSE \cite{feng2020language} as a text encoder, following the same experimental setting as in our paper, under the AudioCaps dataset using the Multilingual Contrastive Language-Audio Pretraining (ML-CLAP), our Audio-English Co-Anchor Contrastive Learning (CACL), and our 1-to-K Contrastive Learnin (KCL) training methods, respectively. The detailed experimental results are shown in Tab. \ref{tab:Audio to Caption Retrieval Results using LaBSE as Text Encoder}, \ref{tab:Caption to Audio Retrieval Results using LaBSE as Text Encoder}, \ref{tab:Results of Mean Rank Variance (MRV) using LaBSE as Text Encoder}, where eng in parentheses indicates the retrieval results of English text, while avg indicates the average retrieval results of all languages. KCL achieved the best retrieval performance, followed by KCL and finally ML-CLAP. The experimental results further illustrate the effectiveness of our scheme in improving recall and mitigating the inconsistency problem on the ML-ATR task.

\begin{table}[h]
\centering
\caption{Audio to Caption Retrieval Results using LaBSE as Text Encoder}
\begin{tabular}{lccc}
\hline
Method & R@1 & R@5 & mAP10 \\
\hline
ML-CLAP (eng) & 51.4 & 81.2 & 31.8 \\
ML-CLAP (eng) & 50.4 & 80.7 & 30.4 \\
CACL (eng) & 55.8 & 84.7 & 33.6 \\
CACL (avg) & 53.0 & 83.0 & 31.8 \\
KCL (eng) & \textbf{58.1} & \textbf{85.7} & \textbf{34.7} \\
KCL (avg) & \textbf{56.5} & \textbf{84.5} & \textbf{34.3} \\
\hline
\end{tabular}
\label{tab:Audio to Caption Retrieval Results using LaBSE as Text Encoder}
\end{table}

\begin{table}[h]
\centering
\caption{Caption to Audio Retrieval Results using LaBSE as Text Encoder}
\begin{tabular}{lccc}
\hline
Method & R@1 & R@5 & mAP10 \\
\hline
ML-CLAP (eng) & 38.6 & 74.7 & 53.8 \\
ML-CLAP (avg) & 37.3 & 72.6 & 52.1 \\
CACL (eng) & 40.7 & 75.6 & 55.4 \\
CACL (avg) & 38.9 & 73.7 & 53.4 \\
KCL (eng) & \textbf{42.1} & \textbf{77.3} & \textbf{56.5} \\
KCL (avg) & \textbf{40.3} & \textbf{75.8} & \textbf{55.0} \\
\hline
\end{tabular}
\label{tab:Caption to Audio Retrieval Results using LaBSE as Text Encoder}
\end{table}

\begin{table}[h]
\centering
\caption{Results of Mean Rank Variance (MRV) using LaBSE as Text Encoder}
\begin{tabular}{lc}
\hline
Method & MRV \\
\hline
ML-CLAP & 14.73 \\
CACL & 11.13 \\
KCL & \textbf{9.46} \\
\hline
\end{tabular}
\label{tab:Results of Mean Rank Variance (MRV) using LaBSE as Text Encoder}
\end{table}

\section{Embedding Space Visualisation}
\label{Appe:Embedding Space}
\begin{figure}[htbp]
    \centering
    \includegraphics[scale=0.22]{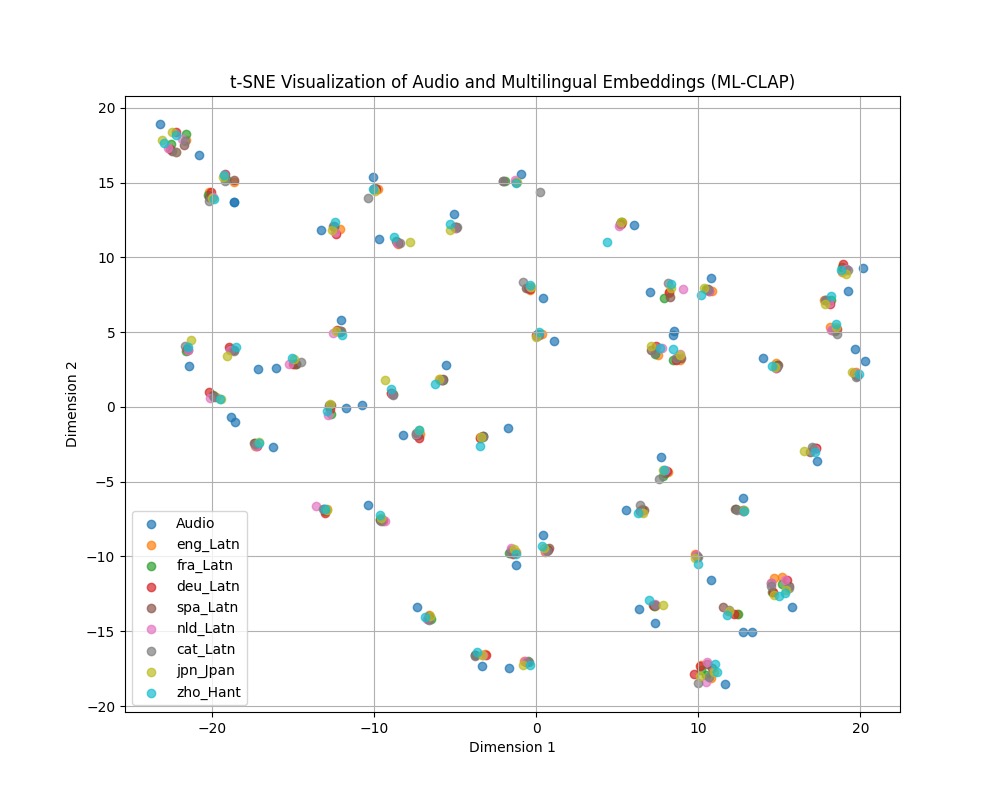}
    \caption{\textbf{Visualisation of the ML-CLAP embedding space}.}
    \label{Fig:Embedding Space ML-CLAP}
\end{figure}

\begin{figure}[htbp]
    \centering
    \includegraphics[scale=0.22]{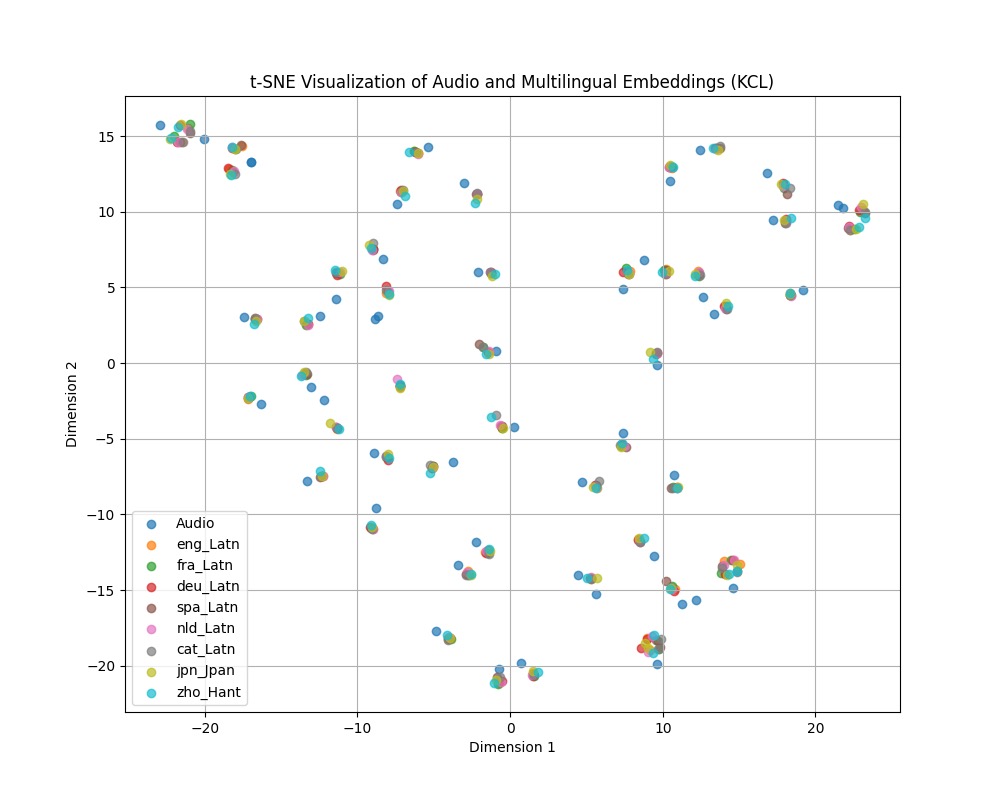}
    \caption{\textbf{Visualisation of the KCL embedding space}.}
    \label{Fig:Embedding Space KCL}
\end{figure}

To further compare the multilingual embedding alignment effects of KCL and ML-CLAP, we randomly select 50 audio-text pairs from AudioCaps. We visualize the embedding spaces of ML-CLAP and KCL after TSNE dimensionality reduction, as shown in Fig. \ref{Fig:Embedding Space ML-CLAP} and Fig. \ref{Fig:Embedding Space KCL}, respectively. In KCL, text embeddings with the same semantics across different languages are more compactly clustered compared to ML-CLAP. This indicates that KCL achieves better alignment of multilingual text embeddings, resulting in more consistent retrieval performance across languages.

\begin{figure}[htbp]
    \centering
    \includegraphics[scale=0.076]{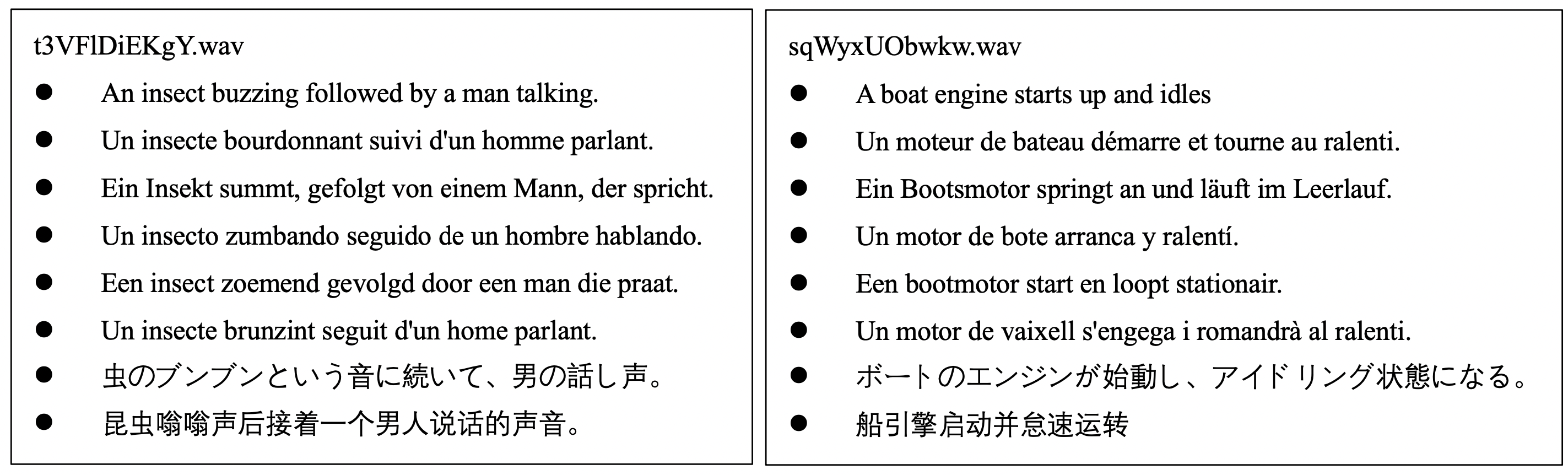}
    \caption{\textbf{Two cases that were successfully retrieved}. The audio text pair on the left is Case 1 and the one on the right is Case 2.}
    \label{Fig:Cases}
\end{figure}

\begin{figure}[htbp]
    \centering
    \includegraphics[scale=0.101]{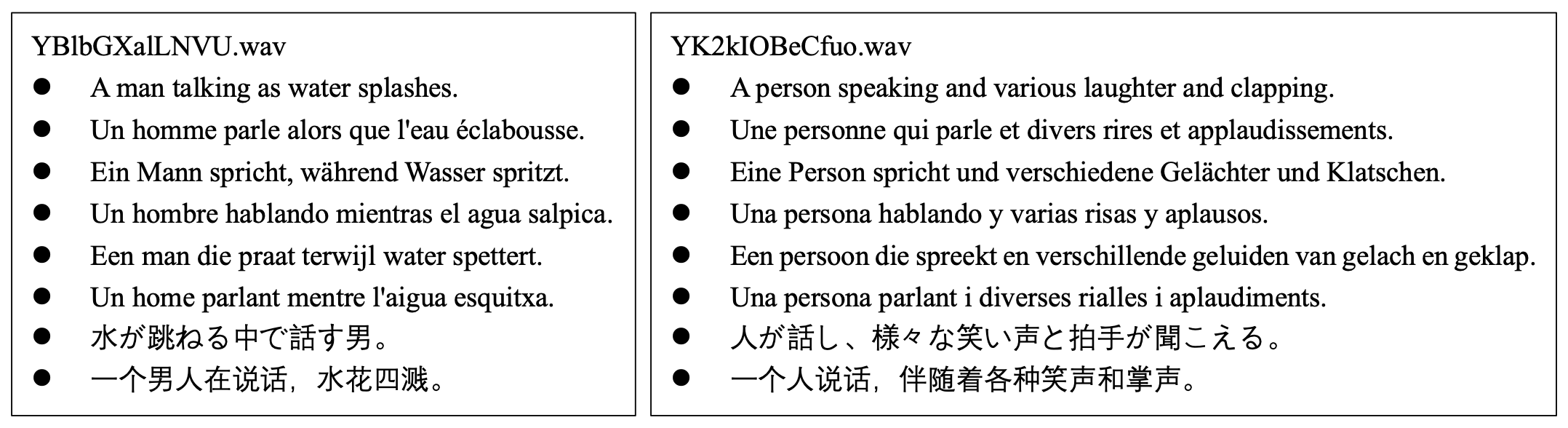}
    \caption{\textbf{Two cases that failed to be retrieved}. The audio text pair on the left is Case 3 and the one on the right is Case 4.}
    \label{Fig:Cases_2}
\end{figure}

\begin{table}[ht]
\caption{Retrieval similarity ranking of successful cases}
\small
\centering
\begin{tabular}{c|cc|cc}
\hline
\multirow{2}{*}{\textbf{Lang}} & \multicolumn{2}{c|}{\textbf{ML-CLAP}} & \multicolumn{2}{c}{\textbf{KCL}}\\ \cline{2-5}
& Case 1 & Case 2 & Case 1 & Case 2 \\ \hline
eng & 5 & 2 & 1 & 0\\ 
fra & 4 & 4 & 1 & 0\\ 
deu & 5 & 4 & 0 & 0\\ 
spa & 2 & 7 & 0 & 0\\ 
nld & 4 & 12 & 0 & 2\\ 
cat & 5 & 12 & 0 & 4\\ 
jpn & 8 & 0 & 0 & 0\\ 
zho & 13 & 3 & 2 & 0\\ \hline
\end{tabular}
\label{Tab:Case Rank}
\end{table}

\begin{table}[ht]
\caption{Retrieval similarity ranking of failed cases}
\small
\centering
\begin{tabular}{c|cc|cc}
\hline
\multirow{2}{*}{\textbf{Lang}} & \multicolumn{2}{c|}{\textbf{ML-CLAP}} & \multicolumn{2}{c}{\textbf{KCL}}\\ \cline{2-5}
& Case 3 & Case 4 & Case 3 & Case 4 \\ \hline
eng & 22 & 14 & 10 & 12\\
fra & 18 & 14 & 8 & 13\\
nld & 33 & 47 & 9 & 17\\
spa & 17 & 20 & 8 & 16\\
deu & 20 & 66 & 7 & 32\\
cat & 24 & 17 & 5 & 14\\
jpn & 11 & 6 & 11 & 2\\
zho & 26 & 8 & 12 & 3\\ \hline
\end{tabular}
\label{Tab:Case Rank 2}
\end{table}

\section{Case Analysis}
\label{Appe:Case Analysis}
In this section, we conduct a cross-lingual retrieval analysis by selecting two groups of representative cases from the AudioCaps test set: two successful retrieval cases and two failed ones. The failed and successful cases are shown in Fig. \ref{Fig:Cases} and \ref{Fig:Cases_2}, respectively. 

Tab. \ref{Tab:Case Rank} and \ref{Tab:Case Rank 2} shows the results of the retrieval rankings of audio-text pairs under the KCL and ML-CLAP schemes. The retrieval rankings of KCL are generally ahead of those of ML-CLAP, and the difference in retrieval rankings across languages is much smaller.

We further analyze the failed cases and found one key commonality: they exhibited excessive semantic overlap with other audio descriptions in the embedding space. This phenomenon is particularly evident in near-identical phrase pairs such as "a person speaks..." versus "a man speaks," where minimal semantic distinction creates challenges for contrastive learning. 

However, even with failed cases, experimental results demonstrate that our proposed KCL method achieves more precise cross-modal alignment between multilingual text and audio embeddings compared to ML-CLAP. This performance advantage substantiates the effectiveness of our approach in handling semantically proximate cases.

\section{Analysis Optimal Alignment Directions for Audio Embedding}
\label{Appe:Analysis Alignment Direction}
Given the audio embedding $\vec a$ and the text embedding vectors of $K$ languages $\{\vec t_1,..., \vec t_K\}$, we need to minimize the total distance between the audio embedding and each language embedding, so that the embedding spaces of different languages are similarly aligned to the audio embedding, and achieve consistent cross-lingual retrieval results. The loss function for minimizing total distance is computed as follows:

\begin{equation}
    \mathcal{L}=\sum^K_{k=1}||\vec a-\vec t_k||^2.
\end{equation}

Derive the loss function $\mathcal{L}$ with respect to embedding $\vec a$:

\begin{equation}
    \frac{\partial L}{\partial\vec a}=\sum^{K}_{k=1}2(\vec a-\vec t_k).
\end{equation}

Making the derivative a zero vector gives:

\begin{equation}
    \sum^K_{k=1}(\vec a-\vec t_k)=0,
\end{equation}

\begin{equation}
    \vec a= \frac{1}{K}\sum^K_{k=1}\vec t_k.
\end{equation}
Therefore, the optimal alignment direction for the audio embedding $\vec a$ is the arithmetic mean of all language text embeddings.

\section{Proof of Weight Error Upper Bound}
\label{Appe:Proof of Weight Error Upper Bound}
We analyze the upper bound on the weighting error heuristically based on the stochastic gradient descent (SGD) optimization algorithm. The following is a detailed theoretical proof of the upper bound on the weighting error in Eq. \eqref{Eq:weight error}.

$Proof$. Based on the definition of the SGD optimization algorithm, we have:
\begin{equation}
\small
    \begin{aligned}
        \mathbf w_{eT}&=\mathbf w_{eT-1}-\eta\sum_{(a,t)}p(a,t)\nabla_{\mathbf w_{eT-1}}\mathbb{E}_{(a,t)}[log\ p(a,t)],\\
        \mathbf w'_{eT}&=\mathbf w'_{eT-1}-\eta\sum_{(a,t)}p_e'(a,t)\nabla_{\mathbf w'_{eT-1}}\mathbb{E}_{(a,t)}[log\ p(a,t)].
    \end{aligned}
\end{equation}

\begin{equation}
\label{Eq:Proof of weight error}
\small
    \begin{aligned}
        &||\mathbf w_{eT}-\mathbf w'_{eT}||\\
        =&||\mathbf w_{eT-1}-\eta\sum_{(a,t)}p(a,t)\nabla_{\mathbf w_{eT-1}}\mathbb{E}_{(a,t)}[log\ p(a,t)]\\
        &-\mathbf w'_{eT-1}+\eta\sum_{(a,t)}p_e'(a,t)\nabla_{\mathbf w'_{eT-1}}\mathbb{E}_{(a,t)}[log\ p(a,t)]||\\
        \leq&^1||\mathbf w_{eT-1}-\mathbf w'_{eT-1}||\\
        &+\eta||\sum_{(a,t)}p_e'(a,t)\nabla_{\mathbf w'_{eT-1}}\mathbb{E}_{(a,t)}[log\ p(a,t)]\\
        &-\sum_{(a,t)}p(a,t)\nabla_{\mathbf w_{eT-1}}\mathbb{E}_{(a,t)}[log\ p(a,t)]||\\
        =&||\mathbf w_{eT-1}-\mathbf w'_{eT-1}||\\
        &+\eta||\sum_{(a,t)}p_e'(a,t)\nabla_{\mathbf w'_{eT-1}}\mathbb{E}_{(a,t)}[log\ p(a,t)]\\
        &-\sum_{(a,t)}p'_e(a,t)\nabla_{\mathbf w_{eT-1}}\mathbb{E}_{(a,t)}[log\ p(a,t)]\\
        &+\sum_{(a,t)}p'_e(a,t)\nabla_{\mathbf w_{eT-1}}\mathbb{E}_{(a,t)}[log\ p(a,t)]\\
        &-\sum_{(a,t)}p(a,t)\nabla_{\mathbf w_{eT-1}}\mathbb{E}_{(a,t)}[log\ p(a,t)]||\\
        \leq^2&||\mathbf w_{eT-1}-\mathbf w'_{eT-1}||\\
        &+\eta||\sum_{(a,t)}p'_e(a,t)(\nabla_{\mathbf w'_{eT-1}}\mathbb{E}_{(a,t)}[log\ p(a,t)]\\
        &-\nabla_{\mathbf w_{eT-1}}\mathbb{E}_{(a,t)}[log\ p(a,t)])||\\
        &+\eta||\sum_{(a,t)}(p'_e(a,t)-p(a,t))\nabla_{\mathbf w_{eT-1}}\mathbb{E}_{(a,t)}[log\ p(a,t)]||\\
        \leq^3&(1+\eta\sum_{(a,t)}p'_e(a,t)\lambda_{(a,t)})||\mathbf w_{eT-1}-\mathbf w'_{eT-1}||\\
        &+\eta g_{max}(\mathbf w_{eT-1})\sum_{(a,t)}||p'_e(a,t)-p(a,t)||.
    \end{aligned}
\end{equation}

The inequality 1 and 2 hold because the Triangle Inequality $|a+b|\leq|a|+|b|$. The inequality 3 holds because

\begin{equation}
\small
    \begin{aligned}
    g_{max}(\mathbf w_{eT-1})=\max_{(a,t)}||\nabla_{\mathbf w_{eT-1}}\mathbb{E}_{(a,t)}[log\ p(a,t)]||,
    \end{aligned}
\end{equation}

and we assume that $\nabla_{\mathbf w'_{eT-1}}\mathbb{E}_{(a,t)}[log\ p(a,t)]$ and $\nabla_{\mathbf w_{eT-1}}\mathbb{E}_{(a,t)}[log\ p(a,t)]$ are $\lambda_{(a,t)}$-Lipschitz, Gradient trimming can be used in the code implementation to a certain extent to reduce the gradient change in the training process, indirectly reduce the excessive growth of Lipschitz constant, as far as possible to meet the Lipschitz continuity condition.

Based on Eq. \eqref{Eq:Proof of weight error}, let 
\begin{equation}
\small
    \begin{aligned}
    a=(1+\eta\sum_{(a,t)}p'_e(a,t)\lambda_{(a,t)}),
    \end{aligned}
\end{equation}
we have

\begin{equation}
\small
    \begin{aligned}
        &||\mathbf w_{eT}-\mathbf w'_{eT}||\\
        \leq&a||\mathbf w_{eT-1}-\mathbf w'_{eT-1}||\\
        &+\eta g_{max}(\mathbf w_{eT-1})\sum_{(a,t)}||p'_e(a,t)-p(a,t)||\\
        \leq&a^2||\mathbf w_{eT-2}-\mathbf w'_{eT-2}||\\
        &+\eta \sum_{(a,t)}||p'_e(a,t)-p(a,t)||\\
        &(g_{max}(\mathbf w_{eT-1})+ag_{max}(\mathbf w_{eT-2}))\\
        \leq&a^{T}||\mathbf w_{(e-1)T}-\mathbf w'_{(e-1)T}||\\
        &+\eta \sum_{(a,t)}||p'_e(a,t)-p(a,t)||(\sum^{T-1}_{j=0} a^jg_{max}(\mathbf w_{eT-1-j}))).
    \end{aligned}
\end{equation}

Thus Eq. \eqref{Eq:weight error} is proved successfully.

\subsection{Migrating to Adam Optimizer}
\label{Appe:Migrating}
We first give the parameter update computation procedure for the Adam optimizer:
\begin{equation}
    g=\sum_{(a,t)}p(a,t)\nabla_{\mathbf w_{eT-1}}\mathbb{E}_{(a,t)}[log\ p(a,t)]\\
\end{equation}

\begin{equation}
    m_{eT}=\beta_1m_{eT-1}+(1-\beta_1)g
\end{equation}

\begin{equation}
    v_{eT}=\beta_2v_{eT-1}+(1-\beta_2)g\circ g
\end{equation}

\begin{equation}
    \hat m_{eT}=\frac{m_{eT}}{1-\beta_1^{eT}}
\end{equation}

\begin{equation}
    \hat v_{eT}=\frac{v_{eT}}{1-\beta_2^{eT}}
\end{equation}

\begin{equation}
    \mathbf w_{eT}=\mathbf w_{eT-1}-\frac{\eta}{\sqrt{\hat v_{eT}}}\hat m_{eT}.
\end{equation}

$m_{eT}$ is the first-order momentum and $v_{eT}$ is the second-order momentum.

We illustrate that the data distribution error also causes weight error in the Adam optimizer by analyzing momentum. The error upper bound of the first-order momentum $m_{eT}$ can be inferred as follows:

\begin{equation}
\small
\label{Eq:Adam error}
\begin{aligned}
&||m_{eT}-m'_{eT}||\\
=&||\beta_1m_{eT-1}-\beta_1m'_{eT-1}\\
&-(1-\beta_1)\sum_{(a,t)}p(a,t)\nabla_{\mathbf w_{eT-1}}\mathbb{E}_{(a,t)}[log\ p(a,t)]\\
&+(1-\beta_1)\sum_{(a,t)}p'_e(a,t)\nabla_{\mathbf w'_{eT-1}}\mathbb{E}_{(a,t)}[log\ p'_e(a,t)])||\\
\leq &||\beta_1m_{eT-1}-\beta_1m'_{eT-1}||\\
&+(1-\beta_1)||\sum_{(a,t)}p(a,t)\nabla_{\mathbf w_{eT-1}}\mathbb{E}_{(a,t)}[log\ p(a,t)]\\
&-\sum_{(a,t)}p'_e(a,t)\nabla_{\mathbf w'_{eT-1}}\mathbb{E}_{(a,t)}[log\ p'_e(a,t)])||\\
\leq&||\beta_1m_{eT-1}-\beta_1m'_{eT-1}||\\
&+(1-\beta_1)||\sum_{(a,t)}p'_e(a,t)(\nabla_{\mathbf w'_{eT-1}}\mathbb{E}_{(a,t)}[log\ p(a,t)]\\
&-\nabla_{\mathbf w_{eT-1}}\mathbb{E}_{(a,t)}[log\ p(a,t)])||\\
&+(1-\beta_1)||\sum_{(a,t)}(p'_e(a,t)-p(a,t))\nabla_{\mathbf w_{eT-1}}\mathbb{E}_{(a,t)}[log\ p(a,t)]||\\
\leq&(1+(1-\beta_1)\sum_{(a,t)}p'_e(a,t)\lambda_{(a,t)})||\mathbf w_{eT-1}-\mathbf w'_{eT-1}||\\
&+(1-\beta_1) g_{max}(\mathbf w_{eT-1})\sum_{(a,t)}||p'_e(a,t)-p(a,t)||.
\end{aligned}
\end{equation}

Eq. \eqref{Eq:Adam error} shows that data distribution error still influences the upper bound on the first-order momentum error in the Adam optimizer. Similarly, the second-order momentum error is also affected by this error. These momentum errors accumulate in the weight errors, which makes our theoretical error upper bounds applicable under the Adam optimizer as well.






\end{document}